\documentclass[12pt]{article}

\usepackage{scicite}
\usepackage{amsmath}

\usepackage{times}

\def\duzomniejsze{<\kern-.7mm<}
\def\duzowieksze{>\kern-.7mm>}

\def\textbf#1{{\bf #1}}
\def\beq{\begin{equation}}
\def\eeq{\end{equation}}
\def\be{\begin{equation}}
\def\ee{\end{equation}}
\def\ben{\begin{eqnarray}}
\def\een{\end{eqnarray}}
\def\beqa{\begin{eqnarray}}
\def\eeqa{\end{eqnarray}}
\def\eea{\end{array}}
\def\bea{\begin{array}}
\newcommand{\bei}{\begin{itemize}}
\newcommand{\eei}{\end{itemize}}
\newcommand{\bee}{\begin{enumerate}}
\newcommand{\eee}{\end{enumerate}}

\newcommand{\tr}{\operatorname{Tr}}

\def\id{{\rm I}}

\def\>{\rangle}
\def\<{\langle}

\newcommand{\ket}[1]{| #1 \rangle}
\newcommand{\bra}[1]{\langle #1 |}
\newcommand{\proj}[1]{\ket{#1}\bra{#1}}

\def\12{{\textstyle \frac{1}{2}}}
\def\s{\,\,\,}

\def\meszero{\psi_0}
\def\mesone{\psi_1}
\newcounter{lastnote}

\title{For quantum information, two wrongs can make a right}

\author
{Jonathan Oppenheim\\
\normalsize{DAMTP, University of Cambridge, CB3 0WA, Cambridge, UK}\\
}
\date{}

\begin{document} 


\baselineskip24pt


\maketitle


 


 

Can you reliably send information down a telegraph wire that 
doesn't always transmit signals correctly? Claude Shannon put
classical information theory on a firm footing when he showed that
you can correct for transmission errors as long as there is some tiny correlation between
what gets sent and what is received.  What's more, Shannon quantified 
how much information could be reliably communicated.  Classical information
theory was intimately entwined with communication from the onset.
The birth of quantum information theory began from an apparently different
direction -- cryptography -- when it was realized that if you can
reliably send someone a system in a quantum state, then you can use those 
states to exchange private messages
that cannot be cracked by even the most powerful computer~\cite{bb84}.
This cannot be done  classically
without meeting beforehand
to exchange a key
that is as long as the private messages you want to send.  
The field of quantum cryptography is now rather
advanced, but surprisingly, we are still wrestling with the corresponding question 
that was so central to classical information theory: How much quantum information can
we reliably send down a noisy channel.  On page 1812 of this issue~\cite{smith2008quantum}, Smith and Yard
have discovered that we may be further from answering this question 
than we think, but that intriguing clues might come from the very place that
initially sparked our interest in quantum information: cryptography.

Classically, a telegraph wire that is so noisy that no information can be reliably
sent through, is useless.  These are called
zero-capacity channels.
But what about the quantum case?
What about a fiber-optic cable which is so noisy it cannot be used to send any quantum
state reliably?  Because our intuition tends to be classical, it was generally believed
that a channel that cannot convey quantum information would also be useless.  
Yet a few years ago,
the Horodecki brothers and I 
found that although these channels cannot be used to send quantum states,
they can
be used to send classical private messages.  
Indeed, one can
classify all states that, if shared over some channel, are private 
\cite{pptkey}. 
What's more, this privacy is verifiable, which means practical
cryptography can be performed over these zero capacity fibers\cite{HHHLO06}.  
The belief that quantum cryptography required being
able to reliably send quantum states turned out to be wrong.  

Now, Smith and Yard, using results from \cite{SSW2008}, 
have shown a remarkable property of these zero-capacity quantum channels that
can send private messages:
They can be combined with another channel
that also has zero-capacity and these two zero-capacity 
channels can be used to
convey quantum information.  It's a bit like
finding out that $0+0=1$. Each channel individually is useless for 
sending quantum
information, but when used together, they can be used to reliably 
send a quantum system
in any state.  

Despite how perplexing this result appears from
a classical perspective,
%
there is a fairly simple way to illustrate it.
setting from that considered by Smith and Yard.
Let us start with the main idea behind cryptography.  Consider two parties, 
Alice and Bob, who can  communicate classically
using a telephone and exchange quantum states -- for example, polarized photons or qubits.  
These are represented by vectors in a linear superposition of two states, $\ket{0}$ and $\ket{1}$,
so that horizontal polarization of a photon is $\ket{0}$, the vertical polarization is $\ket{1}$,
and linear superposition can give rotations to any angle.
Imagine they can succeed in sharing  
a maximally entangled quantum state whose wave function $\ket{\meszero}$
can be represented as:
\beq
\ket{\meszero}=(\ket{00}_{AB}+\ket{11}_{AB})/\sqrt{2}
\eeq
which is in a superposition of them having their qubits in the $\ket{00}$
state, or in the  $\ket{11}$ state, with Alice (A) possessing one of 
the qubits and Bob (B) in possession of the other.  This state is pure, meaning that nothing in the 
external world can be correlated with it.  As a result, Alice and Bob
can measure in the $\ket{0},\ket{1}$ basis
to obtain a string of correlated and secret bits\cite{ekert91} (their
measurement outcomes will be that Alice and Bob each obtain $0$
or they each obtain $1$).  This string  can be used to share a private
message.  Any channel which can be used to share $\ket{\meszero}$
can be used
to share any other state, and is said to have a positive channel capacity.  Likewise, if they
can share the state 
\beq
\ket{\mesone}=(\ket{00}_{AB}-\ket{11}_{AB})/\sqrt{2}
\label{eq:mesone}
\eeq
which is also maximally entangled but has negative phase, then they can also share both a private message, 
and send quantum states.
Now, consider a channel that half of the time
results in $\ket{\meszero}$ being shared, and the other half of the time, results in 
$\ket{\mesone}$ being shared.  One can show that this channel can only send 
classical messages
since the state that is shared can be rewritten as
\beq
\rho=\frac{1}{2}(\ket{00}_{AB}\bra{00}_{AB}+\ket{11}_{AB}\bra{11}_{AB})\s 
\label{eq:cc}
\eeq
which is just a classically correlated state which can be created using only a telephone. 
This channel cannot create entanglement unless $\ket{\meszero}$ is shared
more often than $\ket{\mesone}$ (or visa versa).  
But, to make it more interesting, the channel 
also sends a {\it flag} -- an additional state which labels which of the two maximally
entangled states
has been sent:
\beq
\rho=\frac{1}{2}(\proj{\meszero}\otimes\tau^0_{A'B'}+\proj{\mesone}\otimes\tau^1_{A'B'})\s .
\label{eq:bellwithflags}
\eeq
If Alice and Bob can distinguish the $ \tau^0_{A'B'}$ flag from the   $ \tau^1_{A'B'}$ flag
by performing measurements on them,
then they will know which of the two entangled states they share and they can
then send private messages or quantum states as before.  They can even perform
a correction to the state to convert $\ket{\mesone}$ into $\ket{\meszero}$.  For example, the flags $\tau^0_{A'B'}$ and  $\tau^0_{A'B'}$ could be orthogonal states 
$\ket{0}_{B'}\bra{0}_{B'}$ and $\ket{1}_{B'}\bra{1}_{B'}$ held by Bob alone, and
by measuring them, he will know which maximally entangled state he shares with Alice, and can act a phase operator $\sigma_Z$ in the case they share $\ket{\mesone}$.

As it turns out, there exist flags that can be completely distinguished when Bob holds the entire
flag, yet are arbitrarily difficult to distinguish when Alice and Bob hold different
parts of the flag and must perform measurements on them in separated labs\cite{ew-hiding}. 
In such cases, they will hardly ever know whether they share $\ket{\meszero}$ or
$\ket{\mesone}$, and their ability to send quantum states to each other is
arbitrarily close to zero (one can make it exactly zero by adding small errors).  
Their inability to distinguish the flags, means that from the point of view of
sending quantum states,
they might as well be sharing the state of Equation (\ref{eq:cc}) rather than that
of Equation (\ref{eq:bellwithflags}).
However, this state is still useful for sending private messages, because
Alice and Bob can still just measure as they did before in the  $\ket{0},\ket{1}$ basis to obtain 
a secret key. An eavesdropper may know whether they share 
$\ket{\meszero}$ or $\ket{\mesone}$ but not whether they obtained $\ket{00}$ or 
$\ket{11}$ after measurement.  The eavesdropper may conceivable be holding
the remainder of a pure state which reduces to either Equations (\ref{eq:cc})
or (\ref{eq:bellwithflags}) on Alice and Bob, but the existence of the flags
restricts what she can learn about Alice and Bob's local states in the case
of Equation (\ref{eq:bellwithflags}). 
Thus channels that produce  states 
close to Equation (\ref{eq:bellwithflags}) 
can be used to share private messages, but they cannot be used to
send quantum information -- they have zero quantum capacity.

Now consider another zero-capacity channel, an {\it erasure channel}, that,
with probability $1/2$ lets the quantum state through perfectly, and the rest
of the time it erases the state; the receiver Bob knows an error occurred
because he will be sent the error
state $\ket{e}$.  
Such a channel turns out to be useless by itself for sending quantum information,
but if Alice first uses the previous zero-capacity private channel and then
puts her half of the flag down the erasure channel, then half of the time, Bob can combine Alice's part of the flag that he receives from this channel, 
with the other half that he received from the
zero-capacity private channel.  He can then distinguish the flag.  So half of the time,
he will know whether they share $\ket{\meszero}$ or $\ket{\mesone}$
and he can perform a correction so that they both share the $\ket{\meszero}$ state.
This means that $3/4$ of the time, Alice and Bob share $\ket{\meszero}$ instead of
$\ket{\mesone}$.  This is significantly greater than half the time, and enough to
create entanglement and get a positive channel capacity.  One can verify that the
resulting shared state $\rho_{AB}$ has positive {\it coherent information}~\cite{coherent} 
$I_c(A\rangle B):=S(B)-S(AB)$ with $S(X)$ the von-Neumann entropy $S(X)=-\tr \rho_X\log\rho_X$.
which gives the quantum capacity in the limit of many uses of the channel.

By using both the zero-capacity
private channel, and the zero-capacity erasure channel together, Alice and Bob can always share the
$\ket{\meszero}$ state.  In the case above, the inputs that Alice sends through the
two channels are not even entangled, but only classically correlated.  What's more,
this procedure can be easily generalized.  All cryptographic protocols must
distill the private states of \cite{pptkey}, and the above protocol can 
be adapted to work
for all of them. From a more technical point of
view, Alice sends Bob the {\it shield} through the erasure channel
and when the shield gets through 
Bob can perform {\it untwisting} to yield $\ket{\meszero}$\cite{pptkey}.
I.e. all private channels or protocols which yield a private bit upon measurement, 
can be thought of as sharing states of the form
\beq
\rho=U\proj{\meszero}\otimes\rho_{A'B'}U^\dagger.
\label{eq:twisted}
\eeq
with $U=\proj{0_B}\otimes \id+\proj{1_B}\otimes V_{AB}$ and  $V_{AB}$ a unitary.  
Since this state is the result of Alice and Bob using the private channel, Alice
need only send her share of $\rho_{A'B'}$ (the shield) down the erasure channel, and when
Bob gets it, he can perform $U^\dagger$ to {\it untwist} the state so that they
share $\ket{\meszero}$.

This result raises many questions, not the  least of which is what this work may say about the
quantum capacity.  
We do not know whether the procedure for activating a private channel is optimal, 
whether every channel that has zero-capacity (but
is not classical) can have positive capacity when combined with another zero-capacity
channel, or even whether every such zero-capacity channel is also a private channel.  
Whatever the answers, it is clear is that the structure of quantum
information theory is much richer than most of us ever anticipated.

\noindent {\bf Acknowledgements}: Thanks to Ashley Lebner and Rob Spekkens for helpful discussions on titles and cartoons.



\bibliographystyle{Science}

\begin{thebibliography}{1}

\bibitem{bb84}
C.~Bennett, G.~Brassard, {\it Proc. of IEEE Conference on Computers, Systems
  and Signal Processing\/} (1984), pp. 175--179.

\bibitem{smith2008quantum}
G.~Smith, J.~Yard, {\it Science} {\bf 321}, 1812 (2008)

\bibitem{coherent}
M.~Nielsen, B.~Schumacher {\it Phys. Rev. A.}, {\bf 54}, 2629 (1996)

\bibitem{pptkey}
K.~Horodecki, M.~Horodecki, P.~Horodecki, J.~Oppenheim, {\it Phys. Rev. Lett\/}
  {\bf 94}, 160502 (2005).

\bibitem{HHHLO06}
K.~Horodecki, M.~Horodecki, P.~Horodecki, D.~Leung, J.~Oppenheim, {\it PRL\/}
  {\bf 100}, 110502 (2008). {q}uant-ph/0608195.

\bibitem{SSW2008}
G.~Smith, J.~A. Smolin, A.~Winter  (2008). Quant-ph/0607039.

\bibitem{ekert91}
A.~Ekert, {\it Phys. Rev. Lett\/} {\bf 67}, 661 (1991).

\bibitem{ew-hiding}
T.~Eggeling, R.~Werner, {\it Phys. Rev. Lett.\/} {\bf 89}, 097905 (2002).

\end{thebibliography}



\noindent {\bf Fig. 1.} 
{\bf Quantum blindsight.} ``You appear to be blind in your left eye and blind in
your right eye.  Why you can see with both eyes is beyond me...'' (Figure not included.
Sorry, but its a cartoon of an optometrist examining a cat.)

\end{document}